\documentclass[twocolumn,letter]{jpsj2}

\def\runtitle{Upper critical field of 3K phase in Sr$_2$RuO$_4$}
\def\runauthor{M. Matsumoto, C. Belardinelli and M. Sigrist}

\title{Upper Critical Field of the 3 Kelvin Phase in Sr$_2$RuO$_4$}

\author{
Masashige Matsumoto
\thanks{
Also at Department of Physics, Faculty of Science,
Shizuoka University, 836 Oya, Shizuoka 422--8529;
E-mail address: spmmatu@itp.phys.ethz.ch
},
Cyril Belardinelli and Manfred Sigrist
}

\inst{Theoretische Physik, ETH--H\"onggerberg, CH--8093 Z\"urich, Switzerland}

\recdate{April 3, 2003}

\newcommand{\Sr}{Sr$_2$RuO$_4$}

\abst{
The inhomogeneous 3 Kelvin phase is most likely a 
  superconducting state nucleating at the interface between
  micrometer-sized 
  Ru-metal inclusions and Sr$_2$RuO$_4$ above the bulk onset of
  superconductivity. This filamentary
  superconducting state yields a characteristic temperature dependence
  of the upper critical field which is sublinear, i.e., $ H_{c2} (T)
  \propto (T^* - T)^{\gamma} $ with $ 0.5 \leq \gamma < 1 $ ($ T^* $:
  nucleation temperature). The Ginzburg-Landau theory is used to
  analyze the behavior of the nucleated spin-triplet phase in a field
  and the characteristic features of $ H_{c2} $ observed in the
  experiment are explained based on a two-component order parameter
  in the presence of a filament of enhanced superconductivity with a
  finite width.
}

\kword{
\Sr, upper critical field, filamentary superconductivity,
Ginzburg-Landau theory
}

\begin{document}
\sloppy
\maketitle

\newcommand{\bA}{{\mbox{\boldmath$A$}}}
\newcommand{\bH}{{\mbox{\boldmath$H$}}}
\newcommand{\bD}{{\mbox{\boldmath$D$}}}
\newcommand{\bk}{{\mbox{\boldmath$k$}}}
\newcommand{\bd}{{\mbox{\boldmath$d$}}}
\newcommand{\bn}{{\mbox{\boldmath$n$}}}
\newcommand{\ex}{\eta_x}
\newcommand{\ey}{\eta_y}
\newcommand{\del}{\partial}

Sr$_2$RuO$_4$ plays an exemplary role among unconventional
superconductors as a realization of spin triplet pairing in a
quasi-two-dimensional (2D) Fermi liquid, with some similarities to
superfluid
$^3$He.
\cite{Maeno-1994,Rice-1995,Baskaran}
Experiments provide strong evidence of a superconducting 
state with in-plane
equal-spin pairing
\cite{Ishida}
and violation of time reversal symmetry.
\cite{Luke}
This uniquely identifies the pairing symmetry to be that of a {\it chiral $p$-wave
state}, analogous to the A-phase of $^3$He: $ \bd(\bk) = \Delta_0 
\hat{z}(k_x \pm ik_y ) $.
\cite{Rice-1998,Maeno-2001}
This is a lucky case in many respects. We mention only a few
points here. (1)
Broken time reversal symmetry is responsible for unusual magnetic
properties.
(2) The order parameter consists of {\it two} complex
components $ {\bf \eta} = ( \eta_x , \eta_y ) $, the only case among all
possible triplet pairing states, for tetragonal crystal symmetry, which
correspond otherwise to one-component order parameters. Thus, we may
write 
\begin{equation}
  \bd ({\bf k}) = \hat{z} ( \eta_x k_x + \eta_y k_y).
\end{equation} 
(3) 
It gives rise to unusual vortex physics, including a square vortex
lattice and anomalous low-temperature flux dynamics.
\cite{Riseman,Agterberg}
(4) Chiral gapless subgap quasiparticle states appear at the surface.
\cite{Matsumoto-1999,Mao}

For these properties the important feature is
the degeneracy of the two order parameter
components, which is guaranteed by the tetragonal symmetry. It was
suggested that this degeneracy lifted by symmetry lowering would lead to
two consecutive phase transitions similar to those in the heavy Fermion
superconductor UPt$_3$.
One of the way to realize this is the confinement of the superconductor in a
narrow filament 
which has by geometry a symmetry lower than tetragonal.\cite{OGAWA}  
Such kind of filamentary superconductivity is
most likely realized in an inhomogeneous superconducting phase dubbed
the "3 Kelvin phase" of Sr$_2$RuO$_4$.
\cite{Maeno-1998,Ando}
This phase appears at nearly double the bulk transition
temperature $ T_c = 1.5 $ K and is associated
with the presence of micrometer-sized Ru-metal inclusions in the otherwise
very clean Sr$_2$RuO$_4$. It has been suggested that this phase nucleates
at the interface between the Ru-metal and Sr$_2$RuO$_4$, where the
critical temperature is larger possibly due to a
locally enhanced density of states and modified electron-electron
interactions.
\cite{Sigrist-2001}
In such a case the superconducting state appears at
a temperature $ T^* > T_c $ in a restricted region of lower
symmetry.  This superconducting state
has a single order parameter component and does not violate
time reversal  symmetry. The component corresponds to the $p$-wave
superconducting state with momentum along the interface, i.e., $ {\bf n}
\cdot {\bf \eta} = 0 $ where $ {\bf n} $ is the interface normal vector.
This filamentary phase yields several unusual properties.
Unlike in a conventional $s$-wave superconductor, the transition to the
bulk phase is not merely a matter of percolation, but represents a real
(time reversal) symmetry-breaking  transition. This corresponds to an
additional second-order phase transition.
\cite{Yoshioka}
Moreover, this system may
constitute a complex intrinsically phase frustrated
superconducting network.

An important way of probing the filamentary
phase is the observation of the nucleation in a magnetic field, i.e., the
upper critical field $ H_{c2} $. The confinement to a narrow filament
yields a sublinear temperature dependence
$ H_{c2} (T) \propto (T^* -T)^{\gamma} $ where
$ \gamma =0.5$ is close to $ T^* $ in contrast to the linear behavior for
the bulk $ H_{c2} $. 
\cite{Buzdin,Abrikosov,Sigrist-2001}
In view of experiments showing exponent $ \gamma $ lying between 0.5 and
1 in qualitative agreement with the expectations,
\cite{Yaguchi}
we would like to reanalyze the behavior of $ H_{c2} $ in the
filamentary phase.

Our analysis is based on the Ginzburg-Landau model
of an infinite planar interface, as introduced in ref.
\ref{ref:Sigrist-2001}.  It was suggested that the locally enhanced $ T_c
$ at the interface  is the result of a local lattice distortion
mainly by RuO$_6$-octahedra rotation around the $z$-axis. This gives
rise to reduced 
hopping matrix elements such that the Fermi velocities decrease,
increasing the density of states with an
additional (Stoner) enhancement of the uniform spin fluctuations. The
extension $ s $ of the distortion is of the order $ 100 $ \AA~and,
\cite{Neutron}
thus, is much shorter than
the coherence length $\xi $.  On the other hand, the size of the Ru
inclusions is $ \sim 1 - 10~\mu {\rm m} $,
large compared to the coherence length.
\cite{Maeno-1998,Ando}
Hence we consider here an infinitely
extended interface separating two half spaces. The purpose of
the present study is to analyze the nucleation of the filamentary
unconventional superconducting state in a field. Therefore, we will
further simplify our model. The system is taken to be symmetric
at the interface so that the material on both sides is identical. This
deviation from reality has only a minor influence on the qualitative
behavior of $ H_{c2} $.
Because the specific orientation of the interface is not so essential, we
choose a concrete example in which the normal vector lies in the
$a-b$ plane and points along the $x$-axis. The interface location 
is then defined by $x=0 $.
Our model is given by the
standard Ginzburg-Landau free-energy functional for the above
$p$-wave order parameter $ (\eta_x, \eta_y) $. The local enhancement of
superconductivity at the interface is introduced by an additional
interface term
\cite{Sigrist-2001}:
\begin{eqnarray}
F &=& \int_{-\infty}^\infty d x \Bigl\{
      a ( |\ex|^2 + |\ey|^2 )
      - \xi \sigma \delta(x) ( |\ex|^2 +|\ey|^2 ) \cr
  &+&  K_1 ( |D_x \ex|^2 + |D_y \ey|^2 ) +
       K_2 ( |D_y \ex|^2 + |D_x \ey|^2 ) \cr
 &+& [ K_3 (D_x \ex)^* (D_y \ey)  
     + K_4  (D_y \ex)^* (D_x \ey) + c.c ] \cr
 &+&   K_5 ( |D_z \ex|^2 + |D_z \ey|^2 ) + \frac{1}{8\pi} (\nabla \times \bA)^2
 \Bigr\}.
\label{eqn:1}
\end{eqnarray}
This term is localized to the interface by a delta function, since
the extension $ s $ is small. Although we will see
later that for a higher field this 
extension can be important, we will ignore it for the moment. 
Moreover, $\bD=\nabla+i(2e/\hbar c)\bA$ and $a=(T-T_c)/T_c$,
where $\bA$ is the vector potential and $T_c=1.5$ K
for the bulk \Sr.  
$K_i$ ($i=1,2,3,4~{\rm and}~5$) are coefficients for the gradient terms (
$K_1/3=K_2=K_3=K_4$ for a cylindrical Fermi surface). We define 
$\xi=\sqrt{K_1}$ as the
characteristic length scale of the order parameter in the bulk material.
For the discussion of the nucleation of the superconducting phase,
we can neglect the quartic terms in the Ginzburg-Landau theory.

The variation of $ F $ with respect to $\ex$ and $\ey$ in a zero-field,
leads to two decoupled differential equations in the form of
Schr\"{o}dinger equations:
\begin{equation}
\left[ - K_{1(2)} \del_x^2 - \xi \sigma \delta(x) \right]
\eta_{x(y)} = -a \eta_{x(y)}.
\label{instab-1}
\end{equation}
The ``lowest energy'' solution corresponds to a bound state for an
attractive delta potential (locally enhanced superconductivity)
leading to an eigenvalue $a$ which 
determines the nucleation temperature.
The order parameter components shows an exponential decay towards the bulk
region:
\begin{equation}
\eta_{x(y)} = \exp(-|x|/\xi_{x(y)}), ~~ \mbox{with} ~~
\xi_{x(y)} = 2 K_{1(2)}/ \xi \sigma
\end{equation}
and the transition temperatures for $\ex$ and $\ey$ are
obtained from the eigenvalues $ a = (T^*-T_c)/T_c  > 0 $:
\begin{equation}
T^*_x = T_c(1+\frac{\sigma^2}{4}) \quad \mbox{and} \quad 
T^*_y = T_c(1+\frac{\sigma^2}{4} \frac{K_1}{K_2}).
\end{equation}
Both are enhanced by the $\sigma$ term and $T^*_y > T^*_x$, since $K_1
> K_2$. Hence the first nucleation of superconductivity occurs in the
$ \ey $-component whose nodes lie perpendicular to $ \bn $. 

We study first the case of fields in the $x-y$ plane. The upper
critical field $ H_{c2} $ for the nucleation of superconductivity is
higher for $\bH \perp \bn $ ($ \parallel y $) than for $\bH \parallel \bn $.
For $\bH \parallel y$, the components $\ex$ and $\ey$ remain decoupled
at the nucleation point. $ H_{c2} $ is determined by the instability
of $\ey$. We use $ \bA=(0,0,-Hx)$ leading to the equation
\begin{equation}
\bigl[ - K_2 \del_x^2 - \xi \sigma \delta(x) + \frac{4 K_5 x^2}{l_H^4} \bigr] \ey = -a \ey
\label{eqn:5}
\end{equation}
for $\ey$, where $l_H=\sqrt{c\hbar/(eH)}$ is the magnetic length.
For low fields (long $l_H$),
the harmonic potential term is a weak perturbation to
(\ref{instab-1}), so that $\ey=\exp(-|x|/\xi_y)$ remains a good
approximation. Substituting it into (\ref{eqn:1}) and integrating over
$ x $ we obtain the second-order term
\begin{equation}
F = a\xi_y + \frac{K_2}{\xi_y} -\xi \sigma + \frac{4K_5 \xi_y^3}{2l_H^4},
\label{eqn:6}
\end{equation}
whose zero determines the instability. Thus $ F=0 $ yields $H_{c2}$:
\begin{equation}
H_{c2} =  \frac{c\hbar}{2e} \sqrt{\frac{2}{K_5
    \xi_y^2}}\sqrt{\frac{T^*_y-T}{T_c}} \; ,
\label{eqn:7} 
\end{equation}
as found previously \cite{Buzdin,Abrikosov,Sigrist-2001}.
With increasing field, however, $l_H$ becomes shorter and the harmonic
potential term in (\ref{eqn:5}) becomes a larger correction. Thus 
the extension of $\ey$ along $ \bn $ shrinks due to the decreasing
cyclotron radius.
A good approximation to the ground state of the ``Schr\"{o}dinger
equation'' (\ref{eqn:5}) is obtained by a variational ansatz, which
captures the basic behavior in a simple way:
\begin{equation}
\ey = \exp(-|x|/\xi_y) \exp(-\sqrt{K_5/K_2} x^2/l_H^2).
\label{eqn:8}
\end{equation}
Here $\exp(-\sqrt{K_5/K_2} x^2/l_H^2)$ describes the asymptotic
behavior for distances far from the interface, while the exponential part
gives the correct boundary conditions at the interface. 
Again we substitute $\ey $ into (\ref{eqn:1}) and integrate over $x$, so
that setting $ F =0
$ leads to $ H_{c2} (T) $ (Fig. 1). We may approximate the low-field range,
by a power law $H_{c2} \propto (T_y^*-T)^\gamma$. We find the best
approximation to the variational solution for
$ \gamma = 0.62 $ in the range $ 0.9T^*_y < T < T^*_y = 2.8$ K.
This value compares well with recent experimental findings of
$\gamma=0.7 - 0.75$, which is indeed sublinear
\cite{Yaguchi}. The exact square-root behavior in the
limit of very small fields is experimentally difficult to observe. 
The limitation of this behavior is given by  $ \xi \ll l_H $, i.e., $ H \ll
H^* 
$ with a characteristic field
$ H^* = 2 \sqrt{K_2/K_5} (\Phi_0/2\pi)/\xi^2 \sim 1 {\rm T} $.
Moreover, in Fig. 1 we observe a deviation from our
variational solution for $T < 2$K.
This is partially due to the limited validity of the
Ginzburg-Landau theory which only extends to the region close to $T_y^*$.
Furthermore, the suppression of $ H_{c2} $ is likely related to a
limiting effect (analogous to the paramagnetic
limiting) which has also been observed in the bulk $ H_{c2} $ for
fields in the basal plane. The discussion of this high-field behavior
lies beyond our model and our scope.

\begin{figure}[t]
\begin{center}
\includegraphics[width=7.8cm]{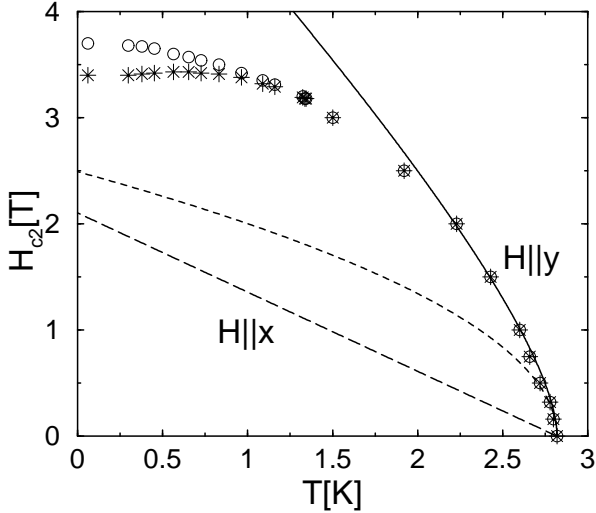}
\end{center}
\caption{
Temperature dependence of the upper critical field for an in-plane field.
The solid line denotes $\bH \parallel y$.
For a low magnetic field,
$H_{c2}$ has a square-root dependence $H_{c2} \propto (T_y^*-T)^{0.5}$,
which is plotted as a dashed line for $\bH \parallel y$.
The long-dash linear line denotes $\bH \parallel x$.
We choose the following parameters:
$K_5/K_1=1/500$ and $K_1/(c\hbar/2e)=\xi^2/(c\hbar/2e)=20$.
$\sigma$ is given to fix the transition temperature $T_y^*=2.8$ K.
Circles (stars) denote the experimental data
for down (up) sweep of field or temperature,
\cite{Yaguchi}
showing a hysteresis in $H_{c2}$ below 1 K.
}
\label{fig:1}
\end{figure}

\begin{figure}[t]
\begin{center}
\includegraphics[width=8.1cm]{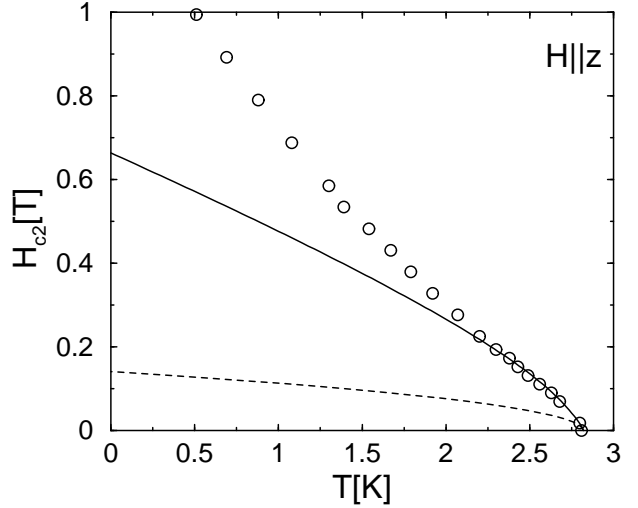}
\end{center}
\caption{
Temperature dependence of the upper critical field for $\bH \parallel z$.
For quite low magnetic fields,
$H_{c2}$ has a square-root dependence, which is plotted as a dashed line.
The parameters are the same as in Fig. \ref{fig:1}.
Circles denote the experimental data.
\cite{Yaguchi}
}
\label{fig:2}
\end{figure}

For $\bH \parallel \bn $, we use $\bA=(0,0,Hy)$ leading to the
Ginzburg-Landau equation for $ \ey $:
\begin{equation}
\bigl[ - K_2 \del_x^2 - K_1 \del_y^2 + \xi \sigma \delta(x)
       + \frac{4 K_5 y^2}{l_H^4} \bigr] \ey = -a \ey.
\label{eqn:9}
\end{equation}
Since the $ x $- and $y$-dependences factorize, we obtain
the relevant solution: 
\begin{equation}
\ey = \exp(-|x|/\xi_y) \exp(-\sqrt{K_5/K_1} y^2/l_H^2)
\; ,
\label{eqn:10}
\end{equation}
and $ H_{c2} \propto T^*_y - T $ as shown in Fig. 1 (long-dash
line). The nucleation field for this direction is obviously
lower. It is clear that the observed $ H_{c2} $ is due to
interfaces that lie parallel to the applied field ($ \bH \perp \bn $),
which yields the highest nucleation field.

Now we turn to fields parallel to the $ z $-axis assuming
simultaneously $ \bH \perp \bn $. In this case the two order
parameter components are coupled. We choose the vector potential as
$\bA=(0,Hx,0)$ so that the free energy is expressed as
\begin{equation} \begin{array}{ll}
F & \displaystyle  =  \int_{-\infty}^\infty dx ( f_x + f_y + f_{xy} ), \\
f_{x(y)} &= a |\eta_{x(y)}|^2 + K_{1(2)} |\del_x \eta_{x(y)}|^2 - \xi
\sigma \delta(x) |\eta_{x(y)}|^2  \\
& \quad \displaystyle 
    + \frac{4 K_{2(1)} x^2}{l_H^4} |\eta_{x(y)}|^2, \\
f_{xy} &= \displaystyle 
\frac{2 (K_3 +K_4)}{l_H^2} \{ i \ex \partial_x \ey^* + i \ey
\partial_x \ex^* +c.c. \} .
\end{array}
\end{equation}
As in the case  of $\bH \parallel y$, we introduce a variational form
for the order parameters:
\begin{equation}
\eta_{y(x)} = C_{y(x)} \exp(-\frac{|x|}{\xi_{y(x)}})
\exp(-\sqrt{K_{1(2)}/K_{2(1)}}\frac{x^2}{l_H^2}) \; ,
\end{equation}
where $C_{x(y)} $ are coefficients to be determined in order to
maximize the nucleation temperature 
for a given magnetic field.
We can integrate the free energy analytically
and determine the upper critical field explicitly.
We plot the result in Fig. \ref{fig:2}.
Our result well reproduces the experiment above $0.9T^*_y$.
In this region, we find a exponent $\gamma=0.76$ fit to a power law,
which agrees well with the experimental result ($\gamma=0.72$).
\cite{Yaguchi}
In the vicinity of $ T^*_y $ $ H_{c2} $ exhibits again a
very-low-field square-root dependence as in the case of $\bH \parallel y$, which is
plotted in Fig.
\ref{fig:2} as a dashed line. In this case, $ H_{c2} $ deviates even more
rapidly from the square-root behavior as temperature decreases than
the in-plane fields. The reason lies in the lower characteristic field
$ H^* = 2 \sqrt{K_2/K_1} (\Phi_0/2\pi) \xi^2 \sim 0.05 {\rm T}$.

\begin{figure}[t]
\begin{center}
\includegraphics[width=8cm]{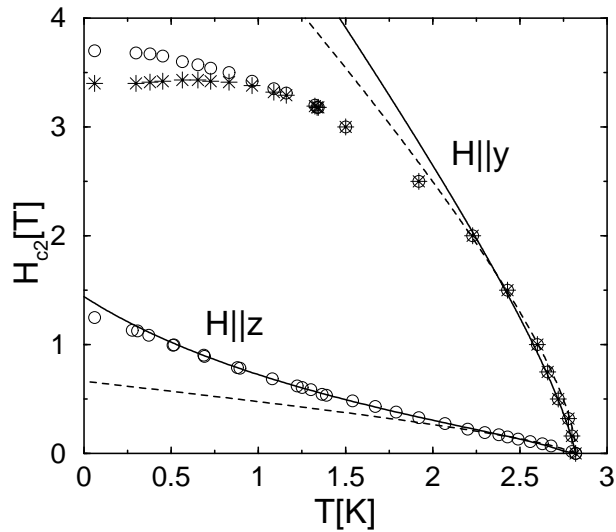}
\end{center}
\caption{
Temperature dependence of the upper critical field for $\bH \parallel y~{\rm and}~z$.
Solid lines are the results for a finite $s$.
Dashed lines represent the result for $s=0$,
which are same as in Figs. \ref{fig:1} and \ref{fig:2}.
We choose the following parameters for a finite $s$:
$s=0.3\xi$, $\lambda=0.7$, $K_5/K_1=1/700$ and $K_1/(c\hbar/2e)=\xi^2/(c\hbar/2e)=24$.
$\sigma$ is fixed to give the transition temperature $T_y^*=2.8$ K.
We used the same experimental data as in Figs. \ref{fig:1} and \ref{fig:2}.
\cite{Yaguchi}
}
\label{fig:3}
\end{figure}

In Fig. 2 the experimental data show a pronounced up-turn
for $T < 2 {\rm K}$,
opposite to the trend for in-plane fields. We would now like to discuss
the origin of this behavior by extending our model. So far we have assumed
that the extension of the region with enhanced superconductivity is
negligible and is well described by a delta function. We replace,
however, now the delta function in eq.(2) by
\begin{equation}
\delta(x) \rightarrow \frac{1}{\sqrt{\pi} s} \exp[-(x/s)^2] \; ,
\end{equation}
where $s$ represents the width of the interface region of
enhanced transition temperature. In the introduction we
speculated that this region is characterized by an increased density
of states or, equivalently, by a decreased Fermi velocity.
Since the coefficient of $K_i$ ($i=1,2,3~{\rm and}~4$)
is connected with the Fermi velocity in the $x-y$ plane ($ K_i \propto
v_F^2/T_c^2 $),
we introduce in addition, the following spatial dependence in $K_i$:
\begin{equation}
K_i \rightarrow K_i \{ 1 - \lambda \exp[-(x/s)^2] \}.~~~(i=1,2,3~{\rm and}~4)
\label{eqn:12}
\end{equation}
with $\lambda$ ($0 \leq \lambda \leq 1$) as a parameter.
Assuming the same variational order parameter form we 
integrate the free energy analytically. The resulting $ H_{c2} $ is
shown in Fig. \ref{fig:3}. In particular, we observe the onset of an
upturn of $ H_{c2} $ for $\bH \parallel z$ at approximately 2 K,
which compares well with the experiment. 

This feature originates from the fact that with shrinking magnetic
length the region of nucleation becomes increasingly confined into a
narrow layer, where we find an enhanced transition temperature as well as a
shorter local coherence length. Both act to increase the critical
field. Thus the onset of the upturn is expected when $ l_H \sim s $.
The fit to the data with our variational approach yields $ s \approx $
200 \AA.
Note that this kind of upturn behavior is not expected for in-plane
fields, since $ K_5 $, that determines the coupling, would not have the
form (\ref{eqn:12}) as it depends on the $z$-axis 
Fermi velocity that would not be significantly affected by RuO$_6$-octahedra
rotations.

In summary, the discussion of the upper critical
field for the filamentary superconducting phase exhibits several
length scales to be taken into account, which
are the effective magnetic length $ l_H $, the
coherence length $
\xi $ and the extension $s$ of the interface regions. If the effective
magnetic length $ l_H $ is much larger than $ \xi $ and $ s $, we find
that $ H_{c2} $ would follow the square-root power law $ (T^*_y - T)^{1/2}
$.  Once the field starts to shrink the extension of the nucleated order
parameter, we encounter a deviation from this behavior and a
power law fit would yield a different power law. Our
discussion shows that in the case of the 3-Kelvin phase the square-root
behavior occurs in a very limited range of low fields which is
hard to analyze. Finally, if the field becomes sufficiently strong to confine
the nucleating order parameter in the interface region $ s $ a
relative increase of the upper critical field is possible. In the
3-Kelvin phase this is observed for the field along the
$z$-axis. However, it has to be noticed that an additional important
enhancement factor is the coupling of the two order parameter
components. This latter effect is due to the Zeeman coupling of the
magnetic field to the orbital magnetic moment of the Cooper pair for the
chiral $p$-wave phase. 

The comparison of our theory with the experiment shows that we are in
principle able to fit the experimental data. However, this aspect has
to be viewed with care, as the Ginzburg-Landau free energy is expanded
at a temperature around $ T^* $ and has therefore limited quantitative
reliability. Moreover the model was in many respects simplified. 
Nevertheless, the ability to reproduce the qualitative 
features are convincing, we believe. One obvious problem
is the limiting behavior for in plane fields which seems to be present in
both the bulk and 3 Kelvin phase of Sr$_2$RuO$_4$.

The upper critical field may be considered a strong confirmation of the
assumption that the 3 K phase \Sr~
is due to the local enhancement of the superconducting transition
temperature at the interface of Ru-inclusions and \Sr~.
Thus, we expect that \Sr~ has two consecutive phase transitions
due to the symmetry lowering by the Ru inclusions. This fact still remains
to be experimentally verified.  It could then be added to the other
convincing evidence for chiral $p$-wave pairing in \Sr.

We are very grateful to Ch. Helm, H. von Loehneysen, Y. Maeno,
H. Monien, 
M. Wada and H. Yaguchi for many valuable discussions and for
providing their experimental data in ref. 19.
This work is financially supported by Japanese Society for the
Promotion of Science (M.M.)
and the MaNEP project of the Swiss National Fund.




\begin{thebibliography}{99}
\bibitem{Maeno-1994}
  Y. Maeno, H. Hashimoto, K. Yoshida, S. Nishzaki, T. Fujita,
  J. G. Bednorz and F. Lichtenberg:
  Nature {\bf 372} (1994) 532.

\bibitem{Rice-1995}
  T. M. Rice and M. Sigrist:
  J. Phys. Condens. Matter {\bf 7} (1995) L643.

\bibitem{Baskaran}
  G. Baskaran:
  Physica B {\bf 223-224} (1996) 490.

\bibitem{Ishida}
  K. Ishida, H. Mukuda, Y. Kitaoka,K. Asayama, Z. Q. Mao, Y. Mori
  and Y. Maeno:
  Nature {\bf 396} (1998) 658.

\bibitem{Luke}
  G. M. Luke, Y. Fudamoto, K. M. Kojima, M. I. Larkin, J. Merrin,
  B. Nachumi, Y. J. Uemura, Y. Maeno, Z. Q. Mao, Y. Mori, H. Nakamura
  and M. Sigrist:
  Nature {\bf 394} (1998) 558.

\bibitem{Rice-1998}
  T. M. Rice:
  Nature {\bf 396} (1998) 627.

\bibitem{Maeno-2001}
  Y. Maeno, T. M. Rice and M. Sigrist:
  Physics Today {\bf 54} (2001) 42.


\bibitem{Riseman}
  T. M. Riseman, P. G. Kealey, E. M. Forgan, A. P. Mackenzie,
  L. M. Galvin, A. W. Tyler, S. L. Lee,
  C. Ager, D. Mck. Paul, C. M. Aegerter, R. Cubitt, Z. Q. Mao,
  S. Akima and Y. Maeno:
  Nature {\bf 396} (1998) 242.

\bibitem{Agterberg}
  D. F. Agterberg:
  Phys. Rev. Lett. {\bf 80} (1998) 5184;
  Phys. Rev. B {\bf 58} (1998) 14484.

\bibitem{Matsumoto-1999}
  M. Matsumoto and M. Sigrist:
  J. Phys. Soc. Jpn. {\bf 68} (1999) 994;
  J. Phys. Soc. Jpn. {\bf 68} (1999) 3120.

\bibitem{Mao}
  Z. Q. Mao, K. D. Nelson, R. Jin, Y. Liu and Y. Maeno:
  Phys. Rev. Lett. {\bf 87} (2001) 037003.



\bibitem{OGAWA}
  N. Ogawa, M. Sigrist and K. Ueda: J. Phys. Soc. Jpn. {\bf 61} (1992) 1730. 

\bibitem{Maeno-1998}
  Y. Maeno, T. Ando, Y. Mori, E. Ohmichi, S. Ikeda, S. NishiZaki and S. Nakatsuji:
  Phys. Rev. Lett. {\bf 81} (1998) 3765.

\bibitem{Ando}
  T. Ando, T. Akima, Y. Mori and Y. Maeno:
  J. Phys. Soc. Jpn. {\bf 68} (1999) 1651.

\bibitem{Sigrist-2001}
  M. Sigrist and H. Monien:
  J. Phys. Soc. Jpn. {\bf 70} (2001) 2409.
  \label{ref:Sigrist-2001}

\bibitem{Yoshioka}
  M. Yoshioka, H. Yaguchi, M. Wada and Y. Maeno:
  Physica C {\bf 388-389} (2003) 501.

\bibitem{Buzdin}
  A. I. Buzdin and L. N. Bulaevski\u{i}:
  Pis'ma Zh. Eksp. Teor. Fiz. {\bf 34} (1981) 118
  [JETP Lett. {\bf 34} (1982) 112].

\bibitem{Abrikosov}
  A. A. Abrikosov: {\it fundamentals of the Theory of Metals},
  (North-Holland, Amsterdam, 1998) Chapter 20.2.

\bibitem{Yaguchi}
  H. Yaguchi, M. Wada, T. Akima, Y. Maeno and T. Ishiguro:
  Phys. Rev. B {\bf 67} (2003) 214519.

\bibitem{Neutron} M. Braden, W. Reichardt, S. Nishizaki, Y. Mori and Y. Maeno:
  Phys. Rev. B {\bf 57} (1998) 1236.


\end{thebibliography}
\end{document}